# Improvement of electronic Governance and mobile Governance in Multilingual Countries with Digital Etymology using Sanskrit Grammar

Arijit Das
Department of Computer Science and Engineering
Jadavpur University
Kolkata, India
arijitdas3@acm.org

Diganta Saha
Department of Computer Science and Engineering
Jadavpur University
Kolkata, India
neruda0101@yahoo.com

*Abstract*— With huge improvement of digital connectivity (Wifi,3G,4G) and digital devices access to internet has reached in the remotest corners now a days. Rural people can easily access web or apps from PDAs, laptops, smartphones etc. This is an opportunity of the Government to reach to the citizen in large number, get their feedback, associate them in policy decision with e governance without deploying huge man, material or resourses.But the Government of multilingual countries face a lot of problem in successful implementation of Government to Citizen (G2C) and Citizen to Government (C2G) governance as the rural people tend and prefer to interact in their native languages. Presenting equal experience over web or app to different language group of speakers is a real challenge. In this research we have sorted out the problems faced by Indo Aryan speaking netizens which is in general also applicable to any language family groups or subgroups.Then we have tried to give probable solutions using Etymology.Etymology is used to correlate the words using their ROOT forms.In 5$^{th}$ century BC Panini wrote Astadhyayi where he depicted sutras or rules- how a word is changed according to person,tense,gender,number etc.Later this book was followed in Western countries also to derive their grammar of comparatively new languages.We have trained our system for automatic root extraction from the surface level or morphed form of words using Panian Gramatical rules.We have tested our system over 10000 bengali Verbs and extracted the root form with 98% accuracy.We are now working to extend the program to successfully lemmatize any words of any language and correlate them by applying those rule sets in Artificial Neural Network.

*Keywords— Digital Etymology; Multilingual eGovernance; Root form extraction; Panini,Neural Network; Semantic Search*

## I. INTRODUCTION

90% of the developing and transit countries are multi lingual that is there are officially more than one languages used for communication between citizens in the same country. .India an example of such countries has has 22 scheduled languages which has official recognition and gets encouragement to promote.Excluding this India has 122 major languages (spoken by more than 10K people) and 1599 other languages. 70% population live in rural area and 90% of them can communicate only in their mother tongue. They feel comfortable to communicate over net in their native language or alternatively it may be said  they interact more over net if they get chance to communicate in their native language.

Naturally it becomes the responsibility and intention of Govt. of multi lingual country to attract those people to e Governance initiative. The constraint of multi linguity can be used as an opportunity here. Govt. has already tried to deploy various websites and applications in front end only in the local languages.But here we have concentrated on a different problem and probable solution.

For C2G governance the most important issue is citizens' feedback.Now a days citizen can give input in their native language (if it is scheduled then normally soft key board is available).But if someone asks any question in Marathi and answer is available in Bengali the system fails to retrieve the answer. In general Search is general keyword matching based.For example if some farmer of Karnataka asking about some problem of coffee farming in Kannada language and the answer or related discussion is already there in Portuguese language as a conversation between two Brazilian farmers. Search engine fails to retrieve the answer as there is no common words and keyword matching algorithm fails.

So technically the problem is as there is a large no. of natural languages to communicate and the system fails to correlate them for search,sort or retrieve. Obviously one of the solution is Machine Translation of all text in other language to English as Internet has the highest data in English language but the problem is there is no dependable automatic machine translator which can convert any language to English with complete accuracy and this is technically an abnormally tedious job to translate each and everything in English and store them online which is nearly impossible.

Fortunately if we look into the anthropology we get Human being first knew how to speak and at the earliest phase those were merely sound signals. Then colloquial languages came. Then formally written alphabets came. Grammar was formed and literature was written but ancient Pandits did not allow the languages to grow haphazardly. That's why there is a huge similarity of Grammar in any language of the world.And historically we get at the early stage of formation of kingdom there was hardly two to three languages officially accepted all over the world.So every language has a root like Sanskrit is the root of all Indo Aryan language.This predicate is supported by the facts all the languages are classified in some group or subgroup and there is also a large inflection between sibling groups.

The first successful solution approach was UNL (Universal Networking Language) which conceptualized and universal language for communication over net but even after 20 years it can't give a universal framework which successfully works. We have taken a different approach Digital Etymology! Etymoogy is the study of the history of words, their origins, and how their form and meaning have changed over time. Which is commonly known as root form extraction in Sanskrit. We have proposed a probable solution to use the knowledge of Paninian Grammar in artificial neural network and measure the central tendency between any two words.

## II. PROBLEM STATEMENT

In all ancient books like the Old Testament or Hindu legends it is mentioned that there was only one language in the world in the beginning era of human civilization. As they spread over the world build different civilization and adopt different languages.

As per the linguists and scientists there are 23 no. of language groups in the world. They are Indo-European, Uralic, Basque, Afro-Asiatic, Niger-Kordofanian, Nilo-Saharan, Khoisan, Altaic, Korean, Japanese, Chukotko-Kamchatkan, Sino-Tibetan, Daic, Austro-Asiatic, Austronesian, Andamanese, Australian, Eskimo-Aleut, Na-Dene, Amerindian, Caucasian, Dravidian and Burushaki.All the languages of a group have originated from one language. Relation between sibling subgrouped languages of same group result huge similarity of surface level words. Now in case of early days of internet communication the medium was mostly English. Later Spanish,Portuguese,French,German,Hindi,Mandarin expanded as medium.But now when internet is reaching in the rural area of developing and transit nation of multilingual countries use of local or native languages is increasing considerably. More it will spread more it will attract untouched native speakers. This is an opportunity for Govt. if language barrier can be overcome and embarrassing if the native speakers get distracted without getting support for their mother tongue over internet.In India for all scheduled languages support for keyboard, codification i.e. support for all insertion and retrieval in front end in native languages have been done. Now the problems are

1. 1. How to correlate same statement expressed in different languages?
2. If they semantically express same thought how to determine and retrieve? (when question and answer are in different languages)
3. 3. How to consider opinions expressed in other languages during decision making? e.g. How to consider Tamil feedbacks when majority are in Hindi only?

These questions are extremely relevant in background processing for proper benefit of eGov.

## III. PROPOSED APPROACH

We are proposing a universal language to represent semantic data extracted from natural language texts as a declarative formal language specifically designed. It can be used as a pivot language in interlingual machine translation systems or as a knowledge representation language in information retrieval applications.

UNL was first such approach to build a global language. We have presented here a probable approach which may be easilyadopted and can overcome the difficulties of UNL.

We first reviewed all the probable language which can be backbone of our proposed universal language. We found German and Sanskrit for using skeletal framework for our proposed language.The reason is

1. Strong grammatical foundation.
2. Rich vocabulary and most importantly
3. The relocation of words in a sentence don't affect the meaning.

And for the last point Sanskrit gives the robust architecture. Wordform(shabd rup) table a 8X3 matrix associates how a word changes according to number,gender, bivakti(to incorporate the meaning of prepositions in the sentence) and root form of the verb (dhaturup) a 3X3 matrix depicts how a verb changes according to tense,person and number.

Any sentence in any language (excluding some rare exception) is either SVO (subject-verb-object) or SOV (subject- object-verb).We propose to encode root form of inflected noun,verb in universal language and special comment field and store them. Therefore a sentence (e.g. I will go) expressed in language A will have same code as of the semantically same sentence (e.g. আমি যাব[ [I will go in bengali]) expressed in language B though there is no common term between them so syntactical search fails to correlate them. Therefore theoretically by this method we can enlarge the domain of reach for an intelligent system to the whole world without any natural language barrier. This gives system an enormous power to search, sort, retrieve, decision making semantically like a human being as oppose to syntactically keyword matching like machines. As the root form are unique and linguists already have done a huge work on etymology, this system will generate a perfect backend process for semantic analysis which can even be extended to paragraphs or text corpus. Taking a simple example "I will go" and "আমি যাব"will have code গমিষ্যামি or root গম-1st person simple future in coded form. Thus both the sentence have same code and though they don't have any common term or syntactically not related but following this method our system can determine that both the system bears same meaning or semantically same. Similarly with the help of "shabdorup" we can encode subject and object of the sentence also and making a composite code for each language.

So it is proved once the sentences are encoded following the above stated ruled it is easy to semantically relate. But now the challenges are
1. How to generate the code in universal language from all natural languages?
2. What will be the mathematical framework for this code generation?

We have started the work and formulated a probable route map to achieve these two goals which have been described in the next section.

IV. METHODOLOGY

We propose a step by step process to generate the code.

1. First determine the Parts of Speech(POS) of each word in a sentence using POS tagger.
2. Mark noun,pronoun and verb and anaphora reference from pronoun to noun.
3. Use supervised learning method specific to each native language to lemmatize them (e.g. running to run)
4. Pass them to complex neural network to determine the Sanskrit root
5. Generate the code
6. Add comment field for handling phrasal verb, idioms etc. to disambiguate the meaning.

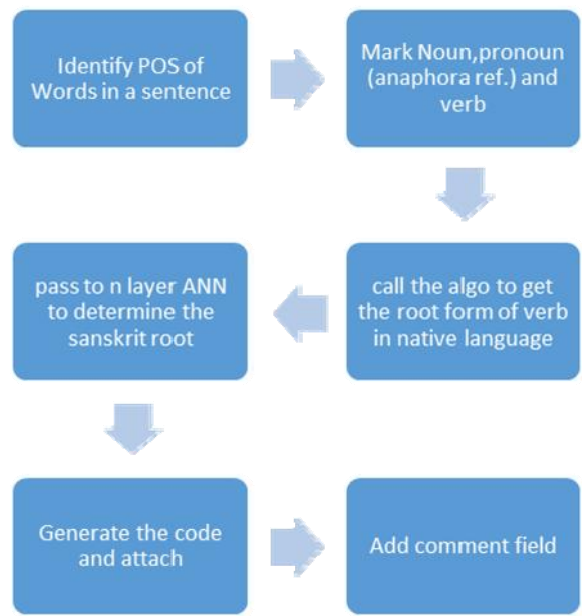

Figure 1. Flowchart of the process

V. WORK DONE

We have already completed the process upto level three taking Bengali as a case study. Our system is giving more than 98% accurate result tested on 10000 verbs.The algorithm of the said process is given as Figure 2.

Our next goal is to fit the set on n level complex neural network to get the Sanskrit root from Bengali root form. The approach is using radix topological sorting of dictionary words as perceptrons and synonyms or matching words (with least levenshtein distance) as sigmoid neurons. Still considerable work pending to announce the success of this experiment and we will publish the same in our next paper.

We have approached to use Parse Thicket to detect anaphora and other relation between words to be incorporated in comment section.

The study of how to improve the system, how to generalize it for all languages and use of semisupervised methods to reduce load on programmer,getting ideas from ongoing research globally are being done.

```
Algorithm 1: DAS & HALDER
Input: Bengali corpus to Shallow parser
       (LTRC)
Output: Inflected verbs are collected in a file
        Input.doc
Input to our system: Input.doc
Output of our system: Multiple files
                      classified according
                      to tense and person
                      consisting of root
                      form of verbs
1  begin
2      Corpus is taken as input to the Shallow
       parser
3      All the inflected verbs are collected in a
       file named as Input.doc
4      Input.doc is taken as input to our system
5      if বিভক্তি(in the Table 1) matches then
6          if person matches then
7              Verbs are passed to the respective
               method with the verbs and বিভক্তি
               for processing
8              Verbs are processed one by one
9              Panini's rules are applied on those
               verbs to extract root verbs
10             Root verb is stored in Output.doc
               file as well as separate file
               according to tense.
11         else
12             The verb collected from Shallow
               parser in Input.doc is not a verb
13     else
14         The verb collected from Shallow
           parser in Input.doc is not a verb
```

Figure 2. Algorithm of step 3 of the process.

## VI. APPLICATION

If successful the universal language will make a revolution in digital communication addressing the need of computing in natural languages.

All the byproduct developed and being developed in each steps are extremely essential in natural language processing research. We have already registered the automatic root verb extractor in ACL.

As there is a large no. of population in multi lingual countries research on this topic will serve the greater no. of people directly and immediately as opposed to incremnental research problems of which outcome comes after a long delay and can benefit very small portion of the society.

It can save the endangered languages by attracting their native speakers to digital media.

Using uniform language in the backend will push us one step ahead for making uniform global village with single language and single nation without the barrier of state, country, race, color thus this will improve the global unity and values of feeling towards united mankind.

This kind of research also enhances cultural exchange heritage values and reduces the narrowness and insulatry.

## VII. SCOPE FOR IMPROVEMENT

As this is a highly interdisciplinary filed of research and specifically we need help from linguists, Sanskrit grammar experts and anthropologists so making a collaborative framework will ease the work.

Government can be directly involved for requirement specification for this universal language at least we need a single language for communication in the backend in this country with population of 120 crores of which dialect changes in every 8 km.

Now a days Govt. is trying to approach to citizens to involve them in governance, this will only be successful if rural people joins and they will join only if they get scope to express their view in mother tongue and that view also should be accountable for decision making thus investing in this research will have a recursive positive effect on e governance, Digital India, social structure, accountability and public administration.